# Black holes to white holes I: A complete quasi-classical model

James M. Bardeen

*Physics Department, Box 1560, University of Washington*
*Seattle, Washington 98195-1560. USA*
*bardeen@uw.edu*

## Abstract

This is the first of two papers presenting different versions of quasi-classical toy models for the non-singular evolution of the geometry and the associated effective stress-energy tensor for a spherically symmetric black hole that evolves into a white hole and eventually disappears after evaporating down to the Planck scale. The ansatz for the geometry is inspired by calculations of the semi-classical stress-energy tensor in the Schwarzschild background and ideas from loop quantum gravity for a nonsingular transition to the white hole. In this paper the main emphasis is on the evolution of the black hole, and the evolution of the white hole is assumed to be essentially the time reverse of that of the black hole. The negative energy of the Hawking "partners" flows out of the white hole to future null infinity. The white hole disappears when the matter and radiation that collapsed to form the black hole emerges. I discuss the compatibility of the model with some of the quantum energy conditions proposed in the literature, and, briefly, the implications for the interpretation of black hole entropy. The second paper considers how the evolution of the white hole can be modified to avoid prolonged emission of negative energy.

## I. INTRODUCTION

The discovery of Hawking radiation from black holes[1] over 40 years ago led to the assertion[2] of a fundamental breakdown of predictability in the evolution of quantum fields following gravitational collapse to form a black hole. The argument was that the Hawking radiation is in a mixed state entangled with negative energy Hawking "partners" inside the black hole that decrease the mass of the black hole to compensate for the positive energy Hawking radiation going to future null infinity. If the black hole has an event horizon and evaporates completely, the result is apparently a loss of quantum information and a breakdown of unitarity for external observers. This "information paradox" is now widely considered an unacceptable conflict with fundamental quantum field theory, requiring drastic departures from the original semi-classical analysis of Hawking, though Unruh and Wald[3] have argued to the contrary. Complete evaporation of the black hole without release of the trapped quantum information does raise serious issues, particularly in the light of the AdS/CFT conjecture[4], in which gravity in the bulk is supposed to be dual to a manifestly unitary conformal field theory on the AdS boundary.

The thermodynamic entropy of a black hole interacting with its surroundings[5] is identified with the Bekenstein-Hawking entropy proportional to the area of the event horizon. In units with $G = c = 1$, $S_{BH} = A/\left(4\hbar\right)$. Normally the thermodynamic entropy of a quantum system is identified with its total number of quantum degrees of freedom, which in turn is the maximum possible value of the entanglement (von Neumann) entropy $S_{vN}$. If the Hawking radiation is entangled with degrees of freedom inside the black hole, as in the standard semi-classical theory of Hawking radiation, Page[6] has shown that $S_{vN}$ becomes equal to $S_{BH}$ at the Page time, when the black hole has lost only about one half of it initial mass. If the black hole continues emitting Hawking radiation after the Page time, as one would expect for any black hole with a mass much greater than the Planck mass $m_p$, either $S_{vN} > S_{BH}$ or the late Hawking radiation must be entangled with the early Hawking radiation. If the latter, by the monogamy of entanglement the late Hawking radiation cannot be entangled with Hawking "partners" inside the black hole horizon, resulting in a "firewall" of highly excited quanta propagating on or just inside the black hole horizon[7].

Controversy over these issues has raged right up to the present time. See reviews by Marolf[8] and Polchinski[9]. A big part of the problem is the lack of a widely accepted theory of quantum gravity. Naively, for very large black holes the semi-classical theory of quantum fluctuations propagating on a classical geometry should be an excellent approximation. Tidal accelerations at the horizon of a very large astrophysical black hole are no larger than those in laboratories on the Earth, where quantum field theory has been tested with exquisite precision. I have argued at length elsewhere[10] that the semi-classical physics in the vicinity of the horizon of a large black hole precludes any substantial storage of quantum information on or near the horizon, and that almost all of the quantum information entangled with the Hawking radiation ends up in the deep interior of the black hole.

However, that does not mean the quantum information is irretrievably swallowed up by a singularity. The classical singularity theorems rely on energy conditions that are violated in quantum field theory. Various more or less ad hoc nonsingular black hole models, some inspired by loop quantum gravity (LQG)[11], have been proposed. One possibility is that quantum backreaction simply stops collapse short of a singularity, which requires an inner trapping horizon. If the inner and outer trapping horizons eventually merge and disappear, the quantum information can escape, as suggested by Hayward[12]. More or less similar models have been proposed by Hossenfelder, et al[13], Rovelli and Vidotto[14], Frolov[15], De Lorenzo, et al[16], and Bardeen[17]. Release of quantum information by the Page time requires a large quantum backreaction in regions of low curvature. The negative surface gravity of the inner trapping horizon raises serious questions about its stability and the viability of these models.

An interesting alternative is the conversion of the black hole into a white hole, as discussed in general terms by Modesto[18] and by Ashtekar and Bojowald[19]. More explicit models are in references [[20],[21],[22],[23],[24]], among others. In some of these there is a Cauchy horizon to the future of the black hole interior, which leaves the unitarity issues unresolved. What is required is a nonsingular quantum transition from the black hole to a white hole jn a spacetime with the causal structure of Minkowski spacetime. The trapped quantum information escapes from the white hole and propagates out to future null infinity. Models that invoke quantum tunneling from a large black hole directly to a white hole, such as that of Haggard and Rovelli[22], I find less convincing than those with a smooth transition of the geometry, as in Ref. [[23]]. Ashtekar, Olmedo, and Singh[24] (AOS) adapted proposals to resolve cosmological singularities in LQG to suggest a particular effective geometry in which the 2-sphere area has a nonzero minimum on a spacelike hypersurface separating the interior of the black hole from the interior of the white hole. However, the AOS model assumes a fixed black hole mass, and has no provision for the Hawking radiation that should dominate quantum corrections at large radii. It is inconsistent with semi-classical quantum theory at large radii where quantum corrections to the geometry are small, as discussed in Part II of this paper.

In Part III I propose a black hole to a white hole model with a smooth effective geometry through the transition, somewhat similar to that of the AOS model while the black hole is large, but that allows the evaporation and eventual disappearance of the black hole at the Planck scale. There are no Cauchy horizons, consistent with unitary evolution for observers at large radii. The model assumes an effective quasi-classical metric even where quantum fluctuations in the geometry are expected to be very large, and therefore should only be considered a suggestion of what might be possible in quantum gravity. The effective stress-energy tensor derived from the effective metric, unlike that of the AOS model, is broadly consistent with the form of the semi-classical stress-energy tensor (SCSET) outside the black hole horizon. A Planck-scale white hole is created (for an external observer) as the black hole disappears, and grows by emitting negative energy.

How the effective stress-energy tensor of the model relates to certain quantum energy conditions is discussed in Part IV. It does seem to satisfy the averaged null energy condition (ANEC) and related quantum null energy condition (QNEC). However, there are contentious issues relating to the evolution of the white hole. I point out reasons to doubt that the claim of De Lorenzo and Perez[25] that instability associated with exponentially increasing blueshifts along the white hole horizon implies a very short lifetime for the white hole. On the other hand, Ref. [[23]] argued for a Planck-scale white hole with a lifetime much longer than that of the black hole. In the model I describe here the evolution of the white hole is roughly the time-reverse of the evolution of the black hole, with the negative energy Hawking "partners" flowing out of the white hole to future null infinity. The prolonged emission of negative energy would seem to violate the Ford-Roman[26] theorems on minimum average energy densities for quantum fields in Minkowski spacetime. Alternatives for the evolution of the white hole will be considered in a companion paper.

Part V has a summary and further discussion of some key issues, such as why the Minkowski minimum energy density theorems may not apply, and why the entanglement entropy of a black hole with a Planck-scale area can greatly exceed its Bekenstein-Hawking entropy.

## II. EFFECTIVE METRICS FROM LQG

Two recent discussions of quantum modifications to the geometry of Schwarzschild black holes, based on slightly different quantization schemes in LQG, are those of Ashtekar and Olmedo[27] (AO), extending the results of AOS to the black hole exterior, and of Gambini, Olmedo, and Pullin[28] (GOP). In AO the square of proper circumferential radius $R$ is expressed in terms of a coordinate $r$ as

$$R^2 = r^2 + \frac{a^2}{4r^2},$$ 
(2.1)

which has a minimum value of $a^2$ at $r^2 = a^2/2$, where there is a smooth transition from trapped surfaces in the black hole $\left(r^2 > a^2/2\right)$ to anti-trapped surfaces in the white hole $\left(r^2 < a^2/2\right)$. AOS argue from LQG that

$$a^2 = \gamma L_0 \delta_c M = \frac{1}{2} \frac{\left(\gamma\right)^{4/3} \Delta^{2/3} M^{2/3}}{\left(4\pi^2\right)^{1/3}},$$ 
(2.2)

Here $\gamma = 0.2375$ is the Barbero-Immirzi parameter of LQG and in terms of the fundamental area gap parameter $\Delta = 5.17\hbar$,

$$L_0 \delta_c = \frac{1}{2} \left(\frac{\gamma \Delta^2}{4\pi^2 M}\right)^{1/3}.$$ 
(2.3)

The mass parameter $M \gg a$ is defined such that $r = 2M$ at the black hole horizon.

Inside the black hole horizon, where $r$ is timelike and the Killing vector $\partial/\partial t$ is spacelike the AO quantum-modified metric can be written as

$$ds^2 = -\left(\frac{R}{r}\right)^2 \frac{\gamma^2 \delta_b^2}{\sin^2\left(\delta_b b\right)} dr^2 + \left(\frac{2M}{R}\right)^2 \frac{\sin^2\left(\delta_b b\right)}{\gamma^2 \delta_b^2} \left[1 + \frac{\sin^2\left(\delta_b b\right)}{\gamma^2 \delta_b^2}\right]^{-2} dt^2$$ 
(2.4)

plus the angular part $R^2 d\Omega^2$, with

$$\cos\left(\delta_b b\right) = b_0 \frac{\left(b_0+1\right)\left(r/2M\right)^{b_0} - \left(b_0-1\right)}{\left(b_0+1\right)\left(r/2M\right)^{b_0} + \left(b_0-1\right)}, \quad b_0 \equiv \sqrt{1 + \gamma^2 \delta_b^2}.$$ 
(2.5)

The coordinate $t$ is singular at $r = 2M$, where $\cos\left(\delta_b b\right) = 1$ and $\sin\left(\delta_b b\right) = 0$, but the continuation to $r > 2M$ is trivial, with

$$\frac{\sin^2 \delta_b b}{\gamma^2 \delta_b^2} \rightarrow -\frac{\sinh^2 \delta_b b}{\gamma^2 \delta_b^2} = \frac{\left[\left(2M/r\right)^{b_0} - 1\right]\left[\left(b_0+1\right)^2 - \left(b_0-1\right)^2\left(2M/r\right)^{b_0}\right]}{\left[b_0 + 1 + \left(b_0-1\right)\left(2M/r\right)^{b_0}\right]^2}.$$ 
(2.6)

The AOS value of $\gamma \delta_b$ is $\gamma \delta_b = 0.5995\left(\hbar/M^2\right)^{1/6}$, and $b_0 - 1 \equiv \varepsilon = 0.1800\left(\hbar/M^2\right)^{1/3}$.

Changing to Eddington-Finkelstein (EF) coordinates, with an advanced time coordinate $v$, constant on ingoing radial null geodesics, resolves the coordinate singularity on the future horizon of the black hole, and the metric becomes

$$ds^2 = -e^{2\psi} g^{rr} dv^2 + 2e^{\psi} dv dr + R^2 d\Omega^2,$$ 
(2.7)

with

$$e^{\psi} = \frac{1}{4}\left(\frac{r}{2M}\right)^{\varepsilon}\left[2+\varepsilon+\varepsilon\left(\frac{2M}{r}\right)^{1+\varepsilon}\right]^2 \cong \left(\frac{r}{2M}\right)^{\varepsilon} \qquad (2.8)$$

and

$$g^{rr} = \left[1-\left(\frac{2M}{r}\right)^{1+\varepsilon}\right]\frac{r^2}{R^2}\frac{\left[(2+\varepsilon)^2-\varepsilon^2\left(2M/r\right)^{1+\varepsilon}\right]}{\left[2+\varepsilon+\varepsilon\left(2M/r\right)^{1+\varepsilon}\right]^2}. \qquad (2.9)$$

This geometry has some very peculiar properties. It is not asymptotically flat in the conventional sense, since $e^{\psi} \to \infty$ as $r \to \infty$. AO argue that by changing the coordinate $t$ to $\tilde{t} = e^{\psi}t$ and taking $r \to \infty$ at constant $\tilde{t}$ the (now non-static) metric does become in a weak sense asymptotically flat, with a well-defined ADM mass. However, the Misner-Sharp quasi-local mass is coordinate-invariant and goes to zero asymptotically. As pointed out by Faraoni and Giusti[29], no initially outgoing timelike geodesics can reach infinite radius.

The quite different effective metric of GOP is based on a LQG spin network with even spacing in circumferential radius $\delta \sim \sqrt{\hbar}$, which is chosen for "simplicity". The effective metric in the coordinates of Eq. (5.3) is

$$ds^2 = -\left[1-\frac{2M}{r}+\frac{\Delta}{4\pi}\frac{(2M)^4}{r^4(r+2M)^2}\right]dv^2 + 2\left(1+\frac{\delta}{2r}\right)dvdr + r^2d\Omega^2. \qquad (2.10)$$

They do not try to model the transition to the white hole, and only consider the effective metric at $r > r_0 \sim \left(\hbar M\right)^{1/3}$. As $r \to \infty$ the energy density falls off as $\delta\left(2M+3\delta/4\right)/r^4$ and the radial and transverse stresses as $\delta/r^3$. The asymptotic Misner-Sharp mass is equal to $M+\delta$. While more reasonable than the AO metric, quantum corrections in the semi-classical regime are still large relative to the semi-classical field theory expectation of quantum corrections proportional to $\hbar$.

Both AOS/AO and GOP completely ignore the evolution of the black hole due to the emission of Hawking radiation. The analyses are based in different ways on symmetry-reduced Hamiltonians, which I expect are inherently incapable of properly accounting for all quantum corrections to the effective metric and stress-energy tensor.

### III. MODELING AN EVAPORATING BLACK HOLE

In constructing a model for the evolution of the geometry of an evaporating black hole and the transition to a white hole, assuming spherical symmetry, it is highly advantageous to work in EF coordinates. The advanced version with advance time $v$ is regular on the black hole trapping horizon. The retarded version, with retarded time $u$ constant on outgoing radial null geodesics, is regular on the white hole anti-trapping horizon. Furthermore, as pointed out by Bardeen[30], the Einstein equations for a general spherically symmetric metric in these coordinates are remarkably simple. The advanced version with the circumferential radius as a coordinate, from now on just denoted by $r$ is

$$ds^2 = -e^{2\psi}g^{rr}dv^2 + 2e^{\psi}dvdr + r^2d\Omega^2, \qquad (3.1)$$

The retarded version differs only in the sign of $g_{ur}$. The Misner-Sharp mass function $m$ has the coordinate-independent definition $\nabla_\alpha r\nabla^\alpha r = 1 - 2m/r$. The Einstein equations that determine the effective stress-energy tensor from $m(v,r)$ and $\psi(v,r)$ are

$$4\pi T_v^{\ v} = -\frac{1}{r^2}\left(\frac{\partial m}{\partial r}\right)_v, \quad 4\pi T_v^{\ r} = \frac{1}{r^2}\left(\frac{\partial m}{\partial v}\right)_r, \quad 4\pi T_r^{\ v} = \frac{1}{r}\left(\frac{\partial e^{-\psi}}{\partial r}\right)_v, \qquad (3.2)$$

with $T_\theta^{\ \theta} = T_\varphi^{\ \varphi}$ from $T_{r;\alpha}^{\ \alpha} \equiv 0$. The retarded version of Eqs. (3.2), with $v \rightarrow u$, is exactly the same. The physical stress-energy tensor components (e.g., $e^{-\psi}T_v^{\ r}$ and $e^{\psi}T_r^{\ v}$) are invariant under $r$-independent rescalings of $v$ (or $u$).

Some models for black hole interiors have suggested a mass function $m$ similar to that of Hayward[12],

$$m = \frac{Mr^3}{r^3 + 2Ma^2}. \qquad (3.3)$$

In the Hayward nonsingular model of an evaporating black hole the mass parameter $M = M(v)$ in the black hole interior, $e^\psi = 1$ in advanced EF coordinates, and $a$ is a constant. The stress-energy tensor curvature invariants are regular at $r = 0$, which is just the origin of a spherical coordinate system in a locally flat geometry. However a non-singular transition to a white hole requires a minimum value of $r > 0$.

I assume a transition to the white hole at a minimum radius $r = a$ similar to that of AOS and AO. While the metric of Eq. (3.1) is singular there, the coordinate singularity can be resolved by changing the radial coordinate from $r$ to $z$ such that

$$r^2 = z^2 + a^2. \qquad (3.4)$$

This is equivalent to Eq. (2.1), but has a simpler form. The coordinate $z$, defined to be negative in the black hole and positive in the white hole, increases to the future inside both the black hole and white hole horizons and is zero at the transition. However, for an *evaporating* black hole, an $a^2$ proportional to $M^{2/3}$, as in AOS and AO, would be time-dependent. Instead, I will make the much simpler assumption that $a^2$ is a Planck-scale constant, perhaps related to the area gap parameter of LQG.

With $z$ instead of $r$ as a coordinate and $a^2$ a constant, the advanced EF metric given in Eq. (3.1) becomes

$$ds^2 = -e^{2\bar\psi_v}g^{zz}dv^2 - 2e^{\bar\psi_v}dvdz + r^2d\Omega^2, \quad g^{zz} = \frac{r^2}{z^2}g^{rr}, \quad e^{\bar\psi_v} = -\frac{z}{r}e^{\psi(v,r)}. \qquad (3.5)$$

The retarded form is

$$ds^2 = -e^{2\bar\psi_u}g^{zz}du^2 - 2e^{\bar\psi_u}dudz + r^2d\Omega^2, \quad g^{zz} = \frac{r^2}{z^2}g^{rr}, \quad e^{\bar\psi_u} = +\frac{z}{r}e^{\psi(u,r)}. \qquad (3.6)$$

In a smooth transition from the black hole to the white hole, $g^{zz}$, $\overline{\psi}_v$, and $\overline{\psi}_u$ vary smoothly, implying $g^{rr} = 1 - 2m/r = 0$ and $e^{\psi} \to \infty$ at $z = 0$. Eqs. (3.2) become

$$4\pi T_v^{\,v} = -\frac{1}{zr}\left(\frac{\partial m}{\partial z}\right)_v, \quad 4\pi T_v^{\,z} = \frac{1}{r^2}\left(\frac{\partial m}{\partial v}\right), \quad 4\pi e^{\overline{\psi}_v}T_z^{\,v} = \frac{a^2}{r^2} - \frac{z}{r^2}\left(\frac{\partial \overline{\psi}_v}{\partial z}\right)_z, \quad (3.7)$$

with

$$2m = r\left(1 - \frac{z^2}{r^2}g^{zz}\right). \quad (3.8)$$

Just replace $v$ by $u$ to get the expressions in retarded EF coordinates.

The causal relationships in my model are illustrated in the Penrose diagram of Fig. 1. The black hole is formed by an influx of matter/radiation along radial null geodesics in a "thick" null shell of mass $M_0$ between advanced times $v_1$ and $v_2$. An infinitesimally thin shell is not physically realistic when considering geometry at close to the Planck scale. The black hole evaporates slowly by emitting Hawking radiation for $0 > v > v_2$, with a "horizon" (*not* an event horizon) defined as the "outgoing" null hypersurface, by definition at $u = 0$, whose radius for $v > v_2$ slowly decreases until trapped surfaces disappear and the black hole ends at the 2-surface where $g^{zz} = 0$ at $r = a$. The "ingoing" null hypersuface at this 2-surface, by definition at $v = 0$, becomes the white hole horizon for $z > 0$. The black hole apparent (trapping) horizon is the timelike hypersurface on which $g^{zz} = 0$ just outside the black hole horizon. The white hole apparent (anti-trapping) horizon is the hypersurface on which , $g^{zz} = 0$, just outside the white hole horizon and timelike if the white hole mass is increasing or just inside the white hole horizon if the white hole mass is decreasing.. In the interior of the collapsing shell there is a spacelike outer trapping horizon indicated by the lower blue line. The white hole ends in the rebounding shell between $u_2$ and $u_1$.

In the model considered in this paper the Hawking "partners" are assumed to propagate along ingoing radial null geodesics in the black hole and along outgoing null geodesics in the white hole and out to future null infinity, as indicated by the black arrows. An alternate picture of partner propagation inside the black hole is propagation along "outward" radial null geodesics (still ingoing in circumferential radius), but this would make little difference in how the black hole transitions to the white hole. If anything, it would even more strongly support the assumption that the Hawking "partners" should propagate along outgoing radial null geodesics in the white hole.

I make no attempt to explicitly model the dynamics of the radiation and evolution of the geometry in the interior of the shell, except to note that inside the shell the geometry should be Minkowski (region M1). However, when the inner edge reaches $r = 0$, quantum backreaction must generate a spacelike inner trapping horizon, indicated by the upper blue line, that connects with the inner edge of the $r = a$ transition hypersurface at the outer edge of the shell, as indicated by the upper blue line. Potential instability due to negative surface gravity of the inner trapping horizon should not be a problem, since it doesn't need to last long. There is another Minkowski region M2 to the future of the rebounding shell.

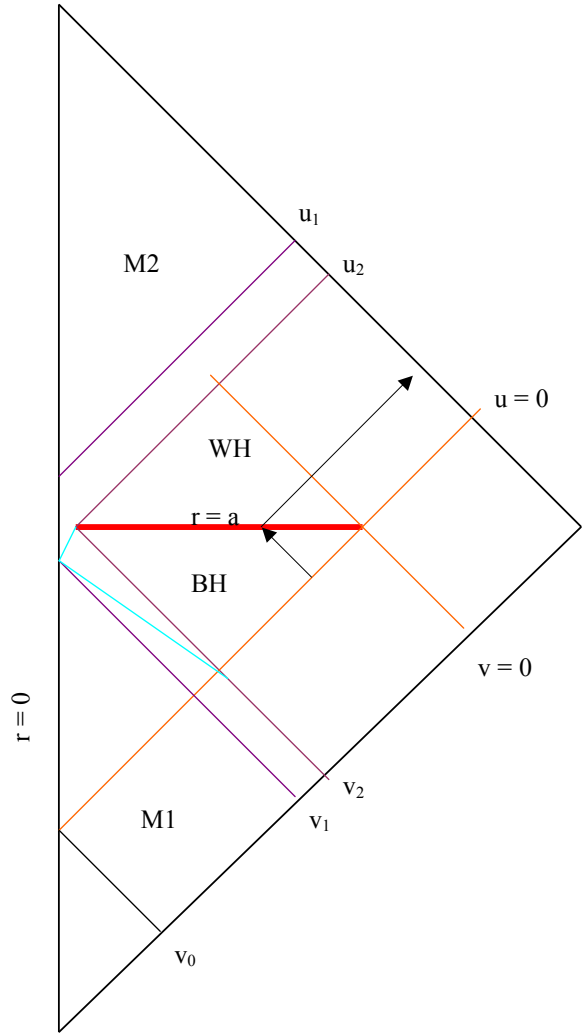

Figure 1. A Penrose diagram showing a thick null shell coming in from past null infinity to form a black hole that transitions into a white hole. See the text for details.

A Penrose diagram can be very misleading as to what events are close to other events. The advanced time $-v_2$ over which the black hole evaporates is $\sim M_0^3 / \hbar$, enormously greater than the range of advanced time $\sim M_0$ over which the black hole forms, and the bounce of the collapsing shell presumably takes place. Also, the Hawking radiation reaches future null infinity over what appears as an infinitesimal range of retarded time in the diagram, but which is actually comparable to $-v_2$ as measured by a distant observers ouside the black hole.

My ansatzes for the metric functions $g^{zz}$ and $e^{\overline{\psi}_v}$ in the region outside the shell are in the spirit of Eq. (3.3), but with added flexibility to better match the form of the SCSET in the exterior of the black hole suggested by numerical calculations for spin 0 and spin 1 fields[10]. Unfortunately, the unknown spin 2 (graviton) contribution to the SCSET presumably dominates, since the spin 2 trace anomaly is more than 10 times the spin 1 trace anomaly in magnitude. Both metric functions should be regular functions of $z^2$ (i.e., $r$) at $z = 0$, implying $1 - 2m/r = 0$ there.

My expression for $g^{\bar{z}\bar{z}}$ is

$$g^{\bar{z}\bar{z}} = 1 - \frac{2Mr^2 + \alpha a^2 r}{r^3 + \beta a^2 r + \gamma (2M) a^2}. \tag{3.9}$$

For a large ($M \gg a$) slowly evaporating black hole, the metric is Schwarzschild in the limit $r \gg \left(Ma^2\right)^{1/3}$ and $M$ is the black hole mass. Calculations of the SCSET show[10] that close to the black hole horizon there is an inflow of negative energy, balancing the outward flow of positive energy Hawking radiation at large radii. Just how this negative energy propagates inside the black hole is somewhat uncertain. I will assume that $M = M(t')$, where $t' = v$ for all $r < 2M$ inside the black hole and should become a retarded time at large radii outside the black hole.

Consider a slowly evaporating black hole, with Hawking luminosity $L_H = -dM/dt \ll 1$. Slow evaporation is plausible throughout the lifetime of the black hole, with $L_H \to 0$ at the end of the black hole. The geometry is quasi-static Schwarzschild for $r > 2M$ as long as long as $2M/a \gg 1$ and $r \ll M/L_H$. For the interpolation between advanced and retarded time, I define $t'$ implicitly by

$$t' = v - 2r + 12M - 16M^2/r - 4M \ln\left(r/2M\right). \tag{3.10}$$

To first order in $L_H$,

$$\left(\frac{\partial t'}{\partial v}\right)_r = \left[1 - 4L_H\left(\ln\frac{r}{2M} - 4 + \frac{8M}{r}\right)\right]^{-1} \tag{3.11}$$

and

$$\left(\frac{\partial t'}{\partial r}\right)_v = \left(-2 - \frac{4M}{r} + \frac{16M^2}{r^2}\right)\left(\frac{\partial t'}{\partial v}\right)_r \tag{3.12}$$

By construction $t'$ and its first derivatives are continuous across $r = 2M$. The ansztz of Eq. (3.10) does not account for Hawking radiation emitted when $M$ and $L_H$ were larger in the past, but this is irrelevant for the current evolution of the black hole. For simplicity, the parameters $\alpha$, $\beta$, and $\gamma$ will just be taken to be constants.

The metric function $\bar{\psi}_v$ controls how the coordinate advanced time $v$ is related to a local proper distances and times. How it varies from one ingoing radial null geodesic to another is a gauge choice, but from Eqs. (3.7) how it varies along these geodesics is related to the $T_{\bar{z}}^v$ component of the stress-energy tensor. The expectation is that gravitational time dilation could become important due to quantum backreaction in the deep interior of the black hole, corresponding to $e^{\bar{\psi}_v}$ becoming small relative to its asymptotic value, which I take to be one. My ansatz for the interior of the black hole, consistent with the form of the semi-classical stress-energy tensor, has three additional parameters $\delta$, $\varepsilon$, and $\phi$:

$$e^{-\bar{\psi}_v} = \left[ 1 + \delta \frac{a^2}{2Mr} + \varepsilon \frac{a^2}{r^2} + \phi \frac{2Ma^2}{r^3} \right]. \tag{3.13}$$

In the exterior, $r > 2M$, the dominant contribution to the stress-energy tensor at large $r$ is the Hawking radiation, and in retarded EF coordinates the dominant component is $T_u^r = -L_H / \left( 4\pi r^2 \right)$. Transforming to advanced EF coordinates,

$$T_v^r = \left( \frac{\partial u}{\partial v} \right)_r T_u^r = T_u^r, \quad T_r^v = \left( \frac{\partial v}{\partial r} \right)_u \left( \frac{\partial u}{\partial r} \right)_v T_u^r = -4T_u^r \left( 1 + O\left( \frac{2M}{r} \right) \right). \tag{3.14}$$

Then at large $r$

$$\left( \frac{\partial \bar{\psi}_v}{\partial r} \right)_v = 4\pi r T_r^v \Rightarrow \bar{\psi}_v = 4L_H \ln\left( \frac{r}{2M} \right) + O\left( L_H \frac{2M}{r} \right). \tag{3.15}$$

This can be accommodated by modifying the ansatz (3.13) for $r > 2M$ to

$$e^{-\bar{\psi}_v} = \left[ 1 - 4L_H \left( \ln\frac{r}{2M} - 4 + \frac{8M}{r} \right) \right]\left[ 1 + \delta \frac{a^2}{2Mr} + \varepsilon \frac{a^2}{r^2} + \phi \frac{2Ma^2}{r^3} \right], \tag{3.16}$$

consistent with continuity of $T_r^v$ at $r = 2M$. The added factor is just $\left( \partial t' / \partial v \right)_r^{-1}$.

After the $z = 0$ transition to the white hole I switch to retarded Eddington-Finkelstein coordinates and the metric of Eq. (3.6), regular at the white hole apparent horizon, and make the simple, but questionable, assumption that the negative energy associated with the Hawking "partners" after flowing at constant $v$ inside the black hole flows out through the white hole at constant retarded time $u$. This means $t' = t'(u)$ for $z > 0$, with $u$ defined in relation to the advanced time for $z < 0$ by $u = -v$ at $z = 0$. Since $t' = v$ in the black hole, in the white hole $t' = -u$. The expression for $g^{zz}$ in Eq. (3.9) remains the same, but with $M = M(u) = M(-t')$. Then

$$\left( \partial M / \partial u \right)_r = -dM / dt' = +L_H, \quad \left( \partial M / \partial r \right)_u = 0. \tag{3.17}$$

Eq. (3.13) for $e^{-\bar{\psi}_u}$ also remains the same. The $z = 0$ hypersurface must be spacelike, $g^{zz} < 0$, with $g^{zz} \to 0$ at the end of the black hole and the beginning of the white hole at $u = v = 0$. The 2-surfaces with $g^{zz} < 0$ on the white hole side are anti-trapped surfaces, with $r$ increasing to the future on "ingoing" as well as outgoing radial null geodesics.

At the evaporation endpoint the mass parameter $M_{min}$ is, from Eq. (3.9),

$$\frac{2M_{min}}{a} = \frac{1 + \beta - \alpha}{1 - \gamma}. \tag{3.18}$$

A modest restriction on the parameters that considerably simplifies the calculations and interpretation of the model is to take $\alpha = \beta + \gamma$, giving $2M_{\min}/a = 1$. Then $g^{zz}$ becomes

$$g^{zz} = \left(1 - \gamma \frac{a^2}{r^2}\right)\left(1 - \frac{2M}{r}\right) \bigg/ \left(1 + \beta \frac{a^2}{r^2} + \gamma \frac{2Ma^2}{r^3}\right). \tag{3.19}$$

A physically sensible model requires $0 < \gamma < 1$, $\beta > -1$. With these restrictions there is only one apparent horizon for the black hole and one for the white hole, both at $r = 2M$ everywhere outside the matter shell. The existence of Hawking radiation requires the existence of a trapping horizon for the black hole, so I assume that the Hawking luminosity $L_{\mathrm{H}}$ smoothly goes to zero as $M \to M_{\min}$ and $v \to 0$.

The mass function $m$ from Eq. (3.8) can be inserted into the first of Eqs. (3.7), using Eq. (3.19) for $g^{zz}$, with the result for $r \le 2M$

$$8\pi T_v^v = \frac{a^2}{r^4}\left\{1 - \left[\begin{array}{l}\left(1 + \dfrac{a^2}{r^2}\right)(\beta + \gamma) + 2\left(\dfrac{2M}{r}\right) \\[2mm] -\dfrac{z^2}{r^2}\left(2\gamma - 3\gamma\dfrac{2M}{r} + g^{zz}\left(2\beta + 3\gamma\dfrac{2M}{r}\right)\right)\end{array}\right]\left[\dfrac{1}{1 + \beta\dfrac{a^2}{r^2} + \gamma\dfrac{2Ma^2}{r^3}}\right]\right\}. \tag{3.20}$$

Also,

$$4\pi T_v^z = 4\pi\left(\frac{r}{z}\right)T_v^r = -L_{\mathrm{H}}\frac{z}{r}\left(\frac{\partial t'}{\partial v}\right)_r \frac{\left[r^4 + \beta r^2 a^2 - \alpha\gamma a^4\right]}{\left[r^3 + \beta ra^2 + \gamma(2M)a^2\right]^2}. \tag{3.21}$$

The vanishing of $T_v^z$ at $z = 0$ is consistent with a smooth transition from inflow of (negative) energy in the black hole to outflow of negative energy at constant $u$ in the white hole.

The $R_z^v = 8\pi T_z^v$ Einstein equation gives inside the black hole

$$4\pi e^{\bar{\psi}_v}T_z^v = \frac{a^2}{r^4}\left\{1 - z^2\frac{\left[\delta/(2Mr) + 2\varepsilon/r^2 + 3\phi(2M/r^3)\right]}{\left[1 + \delta a^2/(2Mr) + \varepsilon a^2/r^2 + \phi(2Ma^2/r^3)\right]}\right\}. \tag{3.22}$$

Then $T_z^z$ can be found from the identity

$$T_z^z = -e^{\bar{\psi}_v}T^{zv} = T_v^v - g^{zz}e^{\bar{\psi}_v}T_z^v. \tag{3.23}$$

After the transition to the white hole, $z$ is positive. I assume that the only change in the expressions for $g^{zz}$ and $e^{-\psi_u}$ from $e^{-\psi_v}$ is that $M = M(u)$. Derivatives of $M$ are evaluated using Eqs. (3.17). The expressions for $T_u^u$, $T_u^z$, and $T_z^u$ are the same as Eq. (3.20), Eq. (3.21), and (3.22). $T_u^z$ like $T_v^z$ is positive.

To further clarify the black hole to white hole transition, project $T_\alpha^\beta$ onto an orthonormal tetrad with future-directed 4-velocity $u^\alpha$ and radial unit vector $n^\alpha$ pointing away from the shell. Where $g^{zz} < 0$ inside the black hole and white hole apparent horizons, and particularly in the vicinity of $z = 0$, it is natural to set $u_v = 0$, so the 4-velocity is orthogonal to a spacelike displacement at constant $z$. Since $u^z > 0$, the remaining components in advanced coordinates are

$$u^v = e^{-\overline{\psi}_v} / \sqrt{-g^{zz}}, \quad u^z = \sqrt{-g^{zz}}, \quad u_z = -1/\sqrt{-g^{zz}}. \tag{3.24}$$

The radial basis vector has $n^v > 0$ so

$$n^v = e^{-\overline{\psi}_v} / \sqrt{-g^{zz}}, \quad n^z = 0, \quad n_v = e^{\overline{\psi}_v} \sqrt{-g^{zz}}, \quad n_z = -1/\sqrt{-g^{zz}}. \tag{3.25}$$

The energy density $E$, the energy flux $F$, and the radial stress $P_r$ are

$$E = -T_z^z - \left(-g^{zz}\right)^{-1} e^{-\overline{\psi}_v} T_v^z = -T_z^z - F, \quad P_r = T_v^v - F. \tag{3.26}$$

In retarded coordinates inside the white hole apparent horizon,

$$u^u = e^{-\overline{\psi}_u} / \sqrt{-g^{zz}}, \quad u^z = \sqrt{-g^{zz}}, \quad u_z = -1/\sqrt{-g^{zz}}, \tag{3.27}$$

$$n^u = -e^{-\overline{\psi}_u} / \sqrt{-g^{zz}}, \quad n^z = 0, \quad n_u = -e^{\overline{\psi}_u} \sqrt{-g^{zz}}, \quad n_z = -1/\sqrt{-g^{zz}}. \tag{3.28}$$

The energy density, energy flux, and radial stress are

$$E = -T_z^z - \left(-g^{zz}\right)^{-1} e^{-\overline{\psi}_u} T_u^z = -T_z^z + F, \quad P_r = T_u^u + F. \tag{3.29}$$

Since $e^{-\overline{\psi}_u} T_u^z$ and $e^{-\overline{\psi}_v} T_v^z$ are identical functions of $|z|$ and the black hole and white hole frames are identical at $z = 0$, the energy flux goes smoothly $\left(C^1\right)$ from positive in the black hole to negative in the white hole.

The energy flux is singular at $g^{zz} = 0$, because the $u_v = 0$ frame is infinitely boosted relative to any local inertial frame. A simple choice of frame valid where $g^{zz} > 0$ is the static frame, defined by $u^z = 0$. Then outside the black hole

$$E = -T_v^v - \left(g^{zz}\right)^{-1} e^{-\overline{\psi}_v} T_v^z = -T_v^v - F, \quad P_r = T_z^z - F. \tag{3.30}$$

Outside the white hole

$$E = -T_u^u - \left(g^{zz}\right)^{-1} e^{-\overline{\psi}_u} T_u^z = -T_u^u + F, \quad P_r = T_z^z + F. \tag{3.31}$$

Taking into account the change in frame across $g^{zz} = 0$, the signs of $F$ and $E + P_r$ do not change across an apparent horizon, $E \cong P_r \cong -F$ for the black hole, and $E \cong P_r \cong +F$ for the white hole. There is no singularity in $F$ in a free-fall frame.

The $G_\theta^\theta$ component of the Einstein tensor is rather complicated, and $T_\theta^\theta = T_\varphi^\varphi$ can most easily be found from the $T_{r;\mu}^\mu = 0$ conservation equation. In advanced coordinates for the black hole,

$$2T_\theta^\theta = \frac{1}{r}\left(r^2 T_z^z\right)_{,r} + re^{-\overline{\psi}_v}\left(e^{\overline{\psi}_v} T_r^v\right)_{,v} - \left(r\overline{\psi}_{v,r} g^{zz} + \frac{r}{2} g^{zz}_{,r}\right) e^{\overline{\psi}_v} T_z^v, \tag{3.32}$$

and similarly for the white hole. $T_\theta^\theta$ is finite at $z = 0$ in spite of a singular term in $T_r^v = \left(r/z\right) T_z^v$, because the singular term does not depend on $v$.

At $z = 0$,

$$g^{zz} = (1-\gamma)(1-2M/a)/(1+\beta+\gamma(2M/a)),$$ (3.33)

The stress-energy tensor reduces to

$$a^2 T_v^v = (-1+2g^{zz})/(8\pi a^2), \quad T_v^z = 0, \quad e^{\overline{\psi}_v} T_z^v = -2(8\pi a^2),$$ (3.34)

The energy density is the same and positive, $E = -T_z^z = +1/(8\pi a^2)$, everywhere on the transition hypersurface. However, $E + P_r = g^{zz}/(4\pi a^2)$ is negative. The expression for $T_\theta^\theta$ is rather complicated in general, but in the limit $2M/a \gg 1$ at $z = 0$, $8\pi a^2 T_\theta^\theta = 10 - 11/\gamma$.

At $r \gg (2Ma^2)^{1/3}$, in the semi-classical regime where quantum corrections to the geometry are small, the SCSET is first-order in $\hbar$, i.e., first-order in an expansion in powers of $a^2$. In this limit

$$2m = 2M + (1+\alpha)\frac{a^2}{r} - \beta\frac{2Ma^2}{r^2} - \gamma\frac{(2M)^2 a^2}{r^3}.$$ (3.35)

The components of the SCSET are polynomials in $x \equiv 2M/r$. Hawking radiation terms only present for $r > 2M$ are enclosed in curly brackets. With $L_H = q(a/2M)^2$,

$$8\pi T_v^v = \frac{a^2}{(2M)^4}\Big[q\{-4-4x+8x^2\}x^2 + (1+\alpha)x^4 - 2(1+\beta)x^5 - 3\gamma x^6\Big],$$ (3.36)

$$8\pi T_v^r = -8\pi T_v^z = -2q\frac{a^2}{(2M)^4}x^2,$$ (3.37)

$$8\pi T_z^v = -2\frac{a^2}{(2M)^4}\Big[q\{4-4x\}x^2 + \delta x^3 + (2\varepsilon-1)x^4 + 3\phi x^5\Big],$$ (3.38)

$$8\pi T_z^z = \frac{a^2}{(2M)^4}\begin{bmatrix}q\{4-20x+16x^2\}x^2 + 2\delta x^3 + (\alpha-2\delta+4\varepsilon-1)x^4 \\ -(2\beta+4\varepsilon-6\phi)x^5 - (3\gamma+6\phi)x^6\end{bmatrix},$$ (3.39)

$$16\pi T_\theta^\theta = \frac{a^2}{(2M)^4}\begin{bmatrix}q\{24x-36x^2\}x^2 - 2\delta x^3 - (2\alpha-5\delta+8\varepsilon-2)x^4 \\ +(6\beta+14\varepsilon-1-18\phi)x^5 + (12\gamma+27\phi)x^6\end{bmatrix}.$$ (3.40)

There is a small discontinuity in $T_\theta^\theta$ at $x = 1$. The trace of the SCSET is

$$8\pi T_\mu^\mu = \frac{a^2}{(2M)^4}\begin{bmatrix}\{-12qx^4\} + (3\delta-4\varepsilon+2)x^4 \\ +(2\beta+10\varepsilon-3-12\phi)x^5 + (6\gamma+21\phi)x^6\end{bmatrix}.$$ (3.41)

Numerical calculations of the Unruh state SCSET in the exterior of a Schwarzschild black hole have been carried out for massless, conformally coupled scalar and vector fields[31] and massless minimally coupled scalar fields[32]. These can be fit[10] within their numerical accuracy by 6th order polynomials in $2M/r$. They all have positive coefficients for the $x^6$ term in $T_\nu^\nu$, corresponding to a negative contribution to $\gamma$, in apparent conflict with my model's requirement that $\gamma > 0$. However, the as yet unknown contribution to the SCSET from spin 2 gravitons should dominate. Additional types if quantum fields will contribute for black holes with masses small compared with those of known astrophysical black holes.

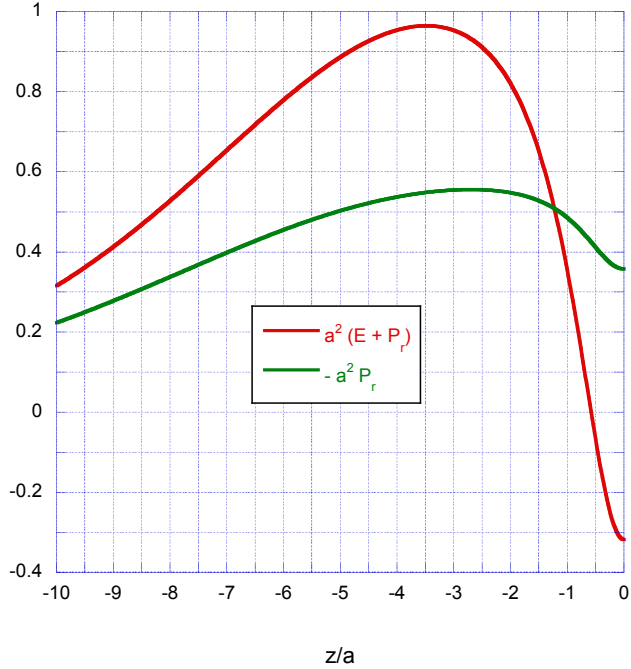

z/a

Figure 2. The energy density and radial stress for the Set A parameters when $2M/a = 8000$ in the core of the black hole for the local frame defined by Eqs. (3.24)-(3.26).

All that is known about the contributions to the SCSET from quantum fluctuations in the gravitational field is the Hawking luminosity and the spin 2 trace anomaly. As the Hawking temperature increases more quantum fields will contribute to the SCSET. If all the quantum fields are conformally coupled, only the $x^6$ term in the trace is nonzero. In the black hole interior, the corresponding constraints from Eq. (3.41) on the coefficients in my model are

$$\varepsilon = 1/2 + 3\delta/4, \quad \beta + 1 + 2\delta - 6\phi = 0. \tag{3.42}$$

A set of model parameters (SetA) consistent with these is

$$\alpha = 0.4, \quad \beta = 0.2, \quad \gamma = 0.2, \quad \delta = 0, \quad \varepsilon = 0.5, \quad \phi = 0.2. \tag{3.43}$$

In the interior of the black hole, with $g^{zz} < 0$ and $F$ negligible, the energy density and radial stress in the orthonormal frame with radial basis vector at constant $z$ are $P_r = T_\nu^\nu$ and $E + P_r = g^{zz} e^{\bar\psi_v} T_z^\nu$ (see Eq. (3.26)). These are plotted in Fig. 2 for the Set A parameters at the advanced time for which $2M/a = 8000$. Quantum modifications to the geometry start becoming unimportant at $z/a \sim -20$.

$E + P_r$ is negative, due to $T_z^v$ becoming positive close to the transition to the white hole, for $-0.6 < z / a < 0.6$.

   Arbitrarily setting $q = 0.001$, about 70 times the value for photons plus gravitons[33] if $a^2 = \hbar$, the energy flux in the core of the black hole for same Set A parameters as in Fig. 2 is plotted in Fig. 3. Even with $q = 1$ the energy flux is smaller than the dominant terms in the stress-energy tensor by a factor of order $(a / 2M)^4$.

   Once the black hole has evaporated down to close to the Planck scale, there is no semi-classical regime inside the horizon and the very notion of a quasi-classical evolution is hard to justify. Still, the model does demonstrate the *possibility* of an evolution in which the black hole ends and the white hole begins without any singularity and without any need for quantum tunneling.

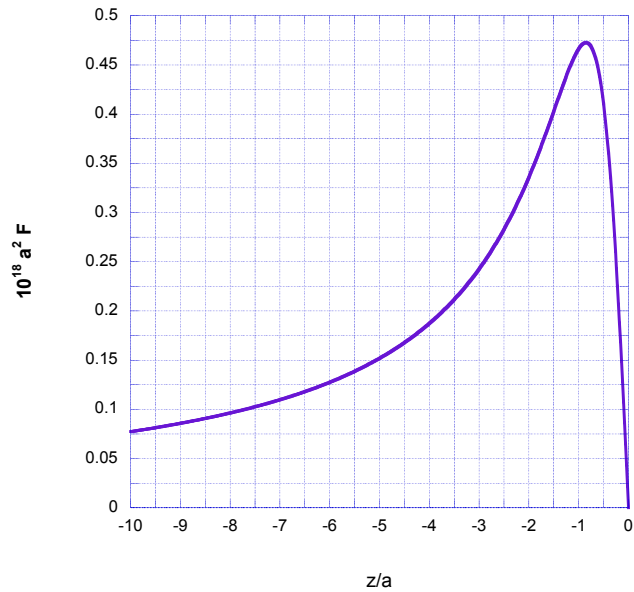

Figure 3. The energy flux $F$ in the core of a black hole for the Set A parameters when $2M / a = 8000$, with $L_H = 0.001(a / 2M)^2$. Compare with Fig. 2, noting that $F$ is smaller than $E$ and $P_r$ by a factor the order of $10^{18}$.

   For what it is worth, I plot in Fig. 4 $E + P_r$ and $P_r$ for the Set A stress-energy tensor for the Set A parameters when $2M / a = 3$. Quantum corrections are significant through the whole black hole interior, and there is no distinct "core" in which the magnitude of the effective stress-energy tensor is slowly varying. The black hole trapping horizon is at $r / a = 3$, $z / a \cong -2.83$. The surface gravity of the trapping horizon at this point is just a bit smaller than the classical value of $1 / 4M$.

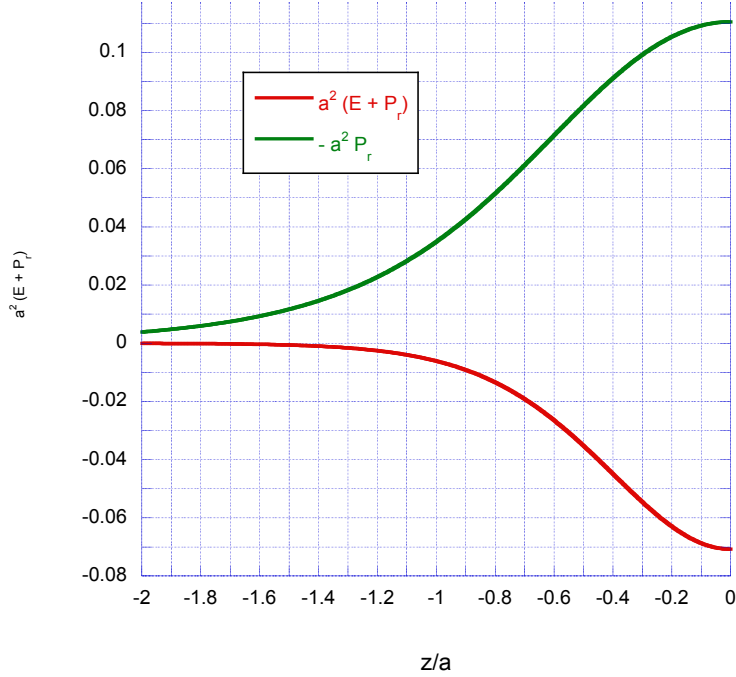

Figure 4. The dominant components of the stress-energy tensor for Set A parameters when $2M/a = 3$. The black hole apparent horizon is at $z/a = -2.83$.
.

## IV. QUANTUM ENERGY CONDITIONS

An interesting question to ask of the model is whether it is consistent with quantum energy conditions that have been proven in some generality in a semi-classical contest. One such condition is the achronal averaged null energy condition[34] (ANEC). This states that

$$\int_{-\infty}^{\infty} T_{\alpha\beta} k^{\alpha} k^{\beta} \, d\lambda \geq 0,$$  (4.1)

where the integral is over a complete *achronal* (no two points connected by a timelike curve) null geodesic with affine parameter $\lambda$ and tangent vector $k^{\alpha} = dx^{\alpha}/d\lambda$. I first consider radial null geodesics crossing the black hole and white hole horizons and then the null generators of the black hole and white hole horizons.

In the black hole region an "ingoing" radial null geodesic has $k^{v} = 0 \Rightarrow k_{z} = 0$ and $k^{z} = -e^{-\bar{\psi}_{v}} k_{v} > 0$, so

$$T_{\alpha\beta} k^{\alpha} k^{\beta} = -e^{-\bar{\psi}_{v}} T_{z}^{v} \left(k_{v}\right)^{2}.$$  (4.2)

From Eq. (3.22) with the Set A parameters, $e^{\bar{\psi}_{v}} T_{z}^{v}$ is negative and slowly varying in most of the core of the black hole, but it must becomes positive close to the transition where $|z/a| < 1$. The evaporation time scale is much longer than a dynamical time scale, and to a good approximation $k_v < 0$ is constant along the trajectory. The integral as the geodesic goes from $z = -\infty$ to $z = 0$ is

$$\int_{-\infty}^{0} T_{\alpha\beta} k^{\alpha} k^{\beta} \left(k^{z}\right)^{-1} dz = +\int_{-\infty}^{0} e^{-\psi_v} \left(e^{\psi_v} T_{z}^{v}\right) k_v \, dz. \tag{4.3}$$

Because $e^{-\psi_v}$ falls off quite rapidly for $r > a$ when $2M/a \gg 1$, the net result for the integral of Eq. (4.3) is typically negative.

As long as $\dot{M}(u) \geq 0$, as assumed, the continuation of the "ingoing" null geodesic into the white hole region stays inside the anti-trapping horizon where $g^{zz} = 0$. The geodesic equation for the ingoing tangent vector in the retarded Eddington-Finkelstein coordinates gives

$$d\left(e^{\psi_u} k^{u}\right)/du = \left(e^{\psi_u} g^{zz}\right)_{,z} \left(e^{\psi_u} k^{u}\right)/2. \tag{4.4}$$

Once $r/a \gg 1$ $e^{\psi_u} k^{u}$ *grows* exponentially, and in the semi-classical regime with a time constant $\cong 4M$. The integral of the null energy can be written as

$$\int \left[-\left(dz/du\right)^2 T_{z}^{u} - T_{u}^{z}\right] e^{\psi_u} k^{u} du. \tag{4.5}$$

During the growth of the white hole, corresponding to the evaporation of the black hole, $\dot{M}(u) > 0$ and $T_{u}^{z} > 0$. The first term is positive except close to $r = a$, but is suppressed as the geodesic approaches the white hole horizon and $dz/du$ becomes very small. The second term in Eq. (4.5) is negative, and while initially small compared to the first term, it quickly becomes dominant. The second term does become positive when the matter and radiation that collapsed to form the black hole starts escaping from the white hole and $\dot{M}(u) < 0$. The exponential growth off $e^{\psi_u} k^{u}$ means that only the last e-folding of the negative contributions is significant, and this is small compared with the positive contribution as the geodesic crosses the matter shell. The same reasoning applies in reverse sequence applies to radially "outgoing" null geodesics passing through the collapsing matter, the interior of the black hole, and exiting across the white hole horizon.

The ANEC is also satisfied for the null generators of the black hole and white hole horizons, since the integrals are also dominated by the positive contributions as they pass through the collapsing matter/radiation shell as the black hole forms and the expanding shell as the white hole disappears.

The quantum null energy condition[35] (QNEC) is a quasi-local lower limit on the null energy based on the von Neumann entropy $S_{vN}$ of the region outside a zero-expansion null hypersurface,

$$\left\langle T_{\alpha\beta} k^{\alpha} k^{\beta} \right\rangle \geq \frac{\hbar}{2\pi A} \frac{d^2 S_{vN}}{d\lambda^2}, \tag{4.6}$$

where $A$ is the area of a cross-section. In the present context, this can be applied at the black hole and white hole horizons, with the cross-section a two-surface of constant $r$ and, respectively, of constant $v$ or $u$. While neither horizon is exactly zero-expansion, they are close enough, at least while $2M/a \gg 1$. Then $S_{vN}(v)$ is slowly increasing on the black hole horizon, and slowly decreasing on the white hole horizon. To a good approximation when $M \gg a$, and with surface gravity $\kappa = 1/\left[4M(v)\right]$ on the black hole horizon,

$$\frac{d^2 S_{vN}}{d\lambda^2} = \frac{d}{d\lambda}\left(\frac{dS_{vN}}{dv}k^v\right) \cong -\kappa\frac{dS_{vN}}{dv}\left(k^v\right)^2, \tag{4.7}$$

since $k^v \propto e^{-\kappa v}$ and

$$\left|\left(d^2 S_{vN}/dv^2\right)/\left(dS_{vN}/dv\right)\right| \ll -\left(dk^v/dv\right)/k^v \cong \kappa. \tag{4.8}$$

On the white hole horizon at $u \cong -v$, $k^u \cong e^{\kappa u}$ and

$$\frac{d^2 S_{vN}}{d\lambda^2} \cong \frac{d}{d\lambda}\left(\frac{dS_{vN}}{du}k^u\right) \cong +\kappa\frac{dS_{vN}}{du}\left(k^u\right)^2, \quad \kappa\frac{dS_{vN}}{du} \cong -\kappa\frac{dS_{vN}}{dv}. \tag{4.9}$$

On both horizons as long as $2M/a \gg 1$,

$$\left\langle T_{\alpha\beta}k^\alpha k^\beta\right\rangle = -\frac{L_H}{4\pi r^2}. \tag{4.10}$$

Using the Hawking luminosity and $dS_{vN}/dv$ as calculated in a semi-classical approximation by Page[36] for photons and gravitons, one can confirm that the QNEC is satisfied while the semi-classical approximation is valid, consistent with the recent claim of a quite general proof of the QNEC in a semi-classical context by Ceyhan and Faulkner[37].

A controversial aspect of the model as formulated in this paper is the negative energy propagating out to future null infinity from the white hole. The asymptotic geometry is Minkowski, and for massless quantum fields in Minkowski spacetime Ford and Roman[27] have established that a lower bound to energy density measured by an inertial observer averaged over a proper time $t_0$ is $E_{min} \sim -m_p^2/t_0^4$. With a mass $M$, at a radius $r$ the time over which tidal accelerations can be neglected means $t_0$ can be as large as $r^{3/2}/M^{1/2}$, corresponding to a minimum averaged energy density $E_{min} \sim -\hbar M^2/r^6$. The negative energy density associated with the negative energy flux from the white hole in my model, falls off roughly as $\hbar/\left(M^2 r^2\right)$, strongly violating the Ford-Roman bound once $r \gg M$. *Any* leakage of negative energy from the white hole lasting much longer than several Planck times has this problem.

Bianchi and Smerlak[38] have made arguments, based on a 2D approximation to black hole evaporation, that an episode of negative energy outflow to future null infinity is *required* in any unitary black hole evaporation scenario. Their result is a necessary condition for unitary evolution of the black hole, in which the von Neumann entropy of the exterior is initially and finally zero,

$$\int_{-\infty}^{\infty} \dot{M}(u)\exp\left[6S_{vN}(u)\right]du = 0. \tag{4.11}$$

This condition is trivially satisfied for my model, but it can also be satisfied by a brief episode of emission of negative energy when the entropy is near its maximum, in this case just after formation of the white hole, that would not violate the Ford-Roman bound.

Prolonged emission of the negative energy accumulated by the black hole to large radii can be avoided, if almost all of it ends up propagating along "ingoing" null geodesics inside or on the white hole horizon, or on timelike geodesics that fall back toward the white hole horizon without reaching large radii. Then the negative energy could eventually be absorbed by the rebounding matter shell that collapsed to form the black hole. Otherwise, the Ford-Roman bound will be violated. Such an alternative scenario for the evolution of the white hole is discussed in the companion paper. Note that the change in sign of the local energy flux from positive in the black hole to negative in the white hole implied by the energy flux in the black hole going to zero at $z = 0$, together with propagation along "ingoing" rather than "outgoing" radial null geodesics implies a *positive* energy density associated with the energy flow as measured by local observers. However, "ingoing" radial null geodesics in the white hole, like "outgoing" null geodesics in the black hole, have negative Killing energy relative to infinity, due to dominance of negative gravitational potential energy, so the contribution to the mass of the white hole is still negative.

While it may seem plausible that the inflow along "ingoing" radial null geodesics in the black hole just continues across the transition to the white hole, this is not necessarily the case. The Hawking "partners" are not point particles following geodesics. They are wave packets of vacuum fluctuations with at least a Planck scale size. A strongly dynamic Planck-scale transition to the white hole can quite plausibly cause a large deviation from geodesic propagation, and convert an "ingoing" null trajectory to an "outgoing" null trajectory.

The advanced EF coordinates in the black hole cannot be continued into the white hole. Consider the equation for an "outgoing" radial null geodesic in the advanced coordinates,

$$\left( \partial z / \partial v \right)_u = -e^{\bar{\psi}_v} g^{zz} / 2. \tag{4.12}$$

Starting from $z$ just greater than zero, $g^{zz}$ is initially negative and becomes positive crossing the white hole apparent horizon. However, at the same point $\left( \partial z / \partial v \right)_u$ must remain positive, which requires that $e^{\bar{\psi}_v} \to \infty$ and change sign at the apparent horizon. What happens at the white hole horizon of the Schwarzschild geometry, with $e^{\psi_v} \equiv 1$, is that $v$ (if defined as here to increase to the future) goes from $+\infty$ to $-\infty$. While in the BH to WH scenario the geometry in the vicinity of the WH horizon well after the transition from the BH may be Schwarzschild to a good approximation, globally the Schwarzschild WH horizon is a Cauchy horizon. The transformation to Kruskal coordinates, which removes the Schwarzschild coordinate singularities, is incompatible with a smooth BH to WH transition.

I will show in the companion paper that a smooth transition from the BH requires initial outflow of negative of negative energy across the WH horizon, but that it is possible to construct scenarios for the evolution of the WH in which this is limited to a relatively short Planck-scale interval of retarded time and does not conflict with the Ford-Roman energy density bound.  Almost all of the negative energy  of the Hawking partners then remains inside a small Planck-scale white hole until the matter/radiation shell emerges.

Finally, the exponentially increasing blueshift of any external energy propagating along the white hole horizon should not be a problem.  There is no reason for a substantial amount of such energy in the context of my model, since the only source for an isolated white hole is the backscatter off of the background curvature of the outgoing Hawking radiation from the black hole and of the outgoing negative energy radiation from the white hole.  The stress-energy tensor of a null fluid is $T^{\alpha\beta} = \sigma k^{\alpha} k^{\beta}$, where $k^{\alpha}$ is a null tangent vector obeying the geodesic equation.  In the retarded coordinates when the geometry is close to Schwarzschild the geodesic equation gives $dk^u / du \cong \left( M / r^2 \right) k^u \cong \kappa k^u$ close to the horizon, with the solution $k^u \cong \left( k^u \right)_0 e^{\kappa u}$.  Then $k^r = -\left( 1 - 2M / r \right) k^u / 2$, from which $r - 2M \cong \left( r - 2M \right)_0 e^{-\kappa u}$, $k_u \cong -\kappa \left( r - 2M \right)_0 \left( k^u \right)_0$ and $k_r \cong -k^u$.  Conservation of the stress-energy gives $d\sigma / du + \sigma k^{\alpha}_{;\alpha} / k^u = 0$.  Since $k^{\alpha}_{;\alpha} = \left( 2 / r \right) \left( dr / du \right) k^u$, $d\sigma / du = \left( r - 2M \right) \sigma / r^2 \propto e^{-\kappa u}$ and $\sigma \to \sigma_0$, a constant.  The contribution to the mass function $m$ from the stress-energy tensor on the horizon is

$$\Delta m \sim -16\pi M^2 \sigma_0 \int k^u k_u \, dr \sim +2\pi M \sigma_0 \left( r - 2M \right)^2 \left( k^u \right)^2, \tag{4.13}$$

which is constant in spite of the exponential blueshift, as is required by energy conservation,  The change in $e^{-\psi_u}$ across the horizon is also unaffected by the blueshift.  Of course, these are classical estimates that do not preclude quantum instabilities.  Actually, the blueshift is locally just an artifact of evaluating the energy in frames accelerating in the opposite direction from the direction of the flow of energy along the horizon.  To the extent that the quantum theory is invariant under local Lorentz transformations, such quantum instabilities should not be present.

In the model presented in this paper, the concern expressed in Ref. [25] that positive energy propagating along the white hole horizon would cause conversion of the white hole into a black hole when it intersects the outgoing shell of rebounding radiation (at $u = u_2$ in Fig 1) is not an issue for the current model, since at that point the backscatter should be predominantly originate from negative energy propagating out of the white hole.

## V. DISCUSSION

At best the toy model of this paper is perhaps representative of the dominant quasi-classical histories contributing to a quantum path integral for evolution of the black hole.  A full quantum gravity treatment is required for any final resolution of the fate of a black hole and the information problem.  The model is not consistent with the existing framework for LQG calculations developed to resolve cosmological singularities. With my choice of parameters, the minimum two-sphere area in the black hole interior is a Planck scale constant perhaps related to the fundamental "area gap" parameter of LQG and is independent of the mass of the black hole. While direct quantum tunneling from the black hole to the white hole at the point the spacetime curvature becomes Planckian, as argued in Ref. [23], might be possible, I would expect the quantum amplitude would be very small compared to that of nonsingular quasi-classical evolution.

I argue that it is reasonable to consider the quantum geometry as small fluctuations about a quasi-classical geometry as long as $r \gg a$, even if this background geometry is substantially modified from a classical solution of the vacuum Einstein equations by quantum backreaction. The effective stress-energy tensor in this quasi-classical geometry is derived from the Einstein tensor calculated from the model metric tensor and is considered to include the macroscopic effects of quantum fluctuations in the gravitational field as well as those of non-gravitational fields. This can make sense as long as individual modes of the quantum fields are small perturbations of a background geometry, even though the cumulative effect of a large number of these modes may substantially modify the geometry. In the context of Schwarzschild, the semi-classical approximation of quantum fields on a fixed classical background geometry should be valid where the spacetime curvature is very sub-Planckian, $M / r^3 \ll m_\mathrm{p}^{-2}$, or $r \gg \left( M m_\mathrm{p}^2 \right)^{1/3}$.

While my guess at the form of the metric in the quasi-classical regime is quite ad hoc, it does match the general form of the SCSET as found by numerical calculations in the literature for spin 0 and spin 1 fields in the Unruh state[30] as extrapolated to the black hole interior, but not necessarily the particular values of the coefficients. The geometry in the model varies smoothly in the transition between the black hole and the white hole throughout the black hole evaporation, even when the black hole horizon area is close to the Planck scale. Of course, one expects large quantum fluctuations in the geometry where $r / a$ is of order one. It would not be surprising if the QNEC were violated there, since it is basically a semi-classical result. The model requires that the quantum focusing conjecture[39] is not valid in the vicinity of the transition to the white hole.

The disturbing feature of this model is that the white hole evolves for most of its lifetime by emitting negative energy. This is the same negative energy that flowed into the black hole during its evaporation. This negative energy must go somewhere. Without prolonged emission of negative energy, the initially Planck scale white hole remains near the Planck scale, and the negative energy is eventually absorbed by the rebounding matter and radiation that formed the black hole.. I will consider this possibility in a companion paper.

Is there some way to rationalize the extended outflow of negative energy from the white hole? The generation of Hawking radiation should be thought of as the tidal disruption of vacuum fluctuations in the vicinity of the black hole horizon, part of which propagate to future null infinity directly with positive energy and part of which end up inside the black hole with negative energy. These parts are not independent of each other. They are strongly entangled and correlated. If the part inside the black hole later propagates out of the white hole to future null infinity, it does not do so as normal "particles", which must have positive energy relative to asymptotic Minkowski vacuum. The negative energy emissions together with the earlier Hawking radiation are still parts of vacuum fluctuations, albeit *very* highly distorted by the black hole geometry.

A somewhat similar situation arises for a zero-energy vacuum fluctuation straddling and propagating along a null hypersurface in Minkowski spacetime. A uniformly accelerating observer for whom that hypersurface is a Rindler horizon becomes infinitesimally close to the horizon in the original inertial frame and only part of the fluctuation is accessible to him. If he eventually stops accelerating, he will gain access to the hidden part of the fluctuation and be able to verify that the energy of the entire fluctuation is zero, but until then the part he can observe may have a small non-zero energy. Important differences from the black hole horizon are no systematic preference in the sign of the energy averaged over many such fluctuations and no conflict the Ford-Roman bound, which applies to inertial observers. The Unruh thermal radiation measured by an accelerating particle detector is not relevant here, since this is a property of the *detector* interacting with the vacuum, and has nothing to do with the stress-energy tensor that is the source in the Einstein equations.

My scenario is incomplete, since there is no explicit modeling of how the collapse of the matter shell is reversed. The $r = a$ minimum radius outside the shell does not apply in its interior, a since at its center $r = 0$ is just the origin of the spherical coordinates in a locally flat region, assuming the bounce can occur without a curvature singularity. What is depicted in Fig. 1 is nothing more than a crude and very schematic guess.

If the black hole does evaporate down to the Planck scale, with no significant release of quantum information across the black hole horizon, as I assume, it is apparent that the Bekenstein-Hawking entropy[5] $S_{\mathrm{BH}} = A / \left( 4\hbar \right) = 4\pi \left( M / m_{\mathrm{p}} \right)^2$ should *not* be interpreted as a measure of the total number of quantum degrees of freedom associated with the black hole. The "partners" of the Hawking radiation quanta simply cross from the black hole region to the white hole region as in Fig. 1 and then flow outward across the white hole horizon. Near the end of the black hole evaporation $S_{\mathrm{BH}}$ is tiny compared with the entropy of the Hawking radiation and the von Neumann entropy of the black hole exterior. It is a mistake to think of the black hole interior degrees of freedom as being in any kind of thermal equilibrium. The degrees of freedom of the bouncing shell and entangled vacuum modes crossing the $z = 0$ spacelike hypersurface are completely out of causal contact with the horizon degrees of freedom of the late stages of the black hole evaporation. While $S_{\mathrm{BH}}$ is presumably a measure of the maximum number of quantum degrees of freedom associated with the black hole horizon at any one time, quantum fluctuations on the horizon do not stay on the horizon. They end up partially in the Hawking radiation and partially after falling deep inside the black hole in what emerges from the white hole. Similar views have been expressed by Garfinkle[40] and Rovelli[41]. This contradicts the "central dogma" behind most papers on the black hole information problem, as reviewed recently by Almheiri, et al[42]. The companion paper will have a more extensive discussion of black hole entropy.

Finally, the assumption of spherical symmetry is unrealistic. Any small deviations from spherical symmetry in the collapse that forms the black hole are amplified as the collapse proceeds, and classically the singularity structure of a Kerr black hole with any nonzero angular momentum is timelike, rather than the spacelike singularity of a Schwarzschild black hole. So does the black hole to white hole transition discussed here have any relevance to an even slightly generic black holes? Bianchi and Haggard[43] have made an initial attempt to address this question. They argue that at least the initial breakdown of the semi-classical approximation in black holes is on a spacelike hypersurface for quantum geometries with small nonzero angular momentum from quantum fluctuations. A black hole to white hole transition with the black hole disappearing at a finite advanced time avoids having to deal with a Cauchy horizon and its associated instabilities, as present in the interior of a classical Kerr black hole, and which would potentially make unitarity for external observers impossible.

ACKNOLEDGEMENTS

This original inspiration for this paper came from discussions with Hal Haggard while we were both visiting the Perimeter Institute. Research at the Perimeter Institute is supported by the Government of Canada through the Department Innovation, Science, and Economic Development, and by the Province of Ontario through the Ministry of Research and Innovation. I also thank Amos Ori and Tommaso De Lorenzo for comments on an earlier version.